# High-pressure structural and elastic properties of Tl$_2$O$_3$

O. Gomis[1,a)] D. Santamaría-Pérez[2,3], J. Ruiz-Fuertes[2,4], J. A. Sans[5], R. Vilaplana[1], H. M. Ortiz[5,6†], B. García-Domene[2], F. J. Manjón[5], D. Errandonea[2], P. Rodríguez-Hernández[7], A. Muñoz[7], and M. Mollar[5]

[1] Centro de Tecnologías Físicas, MALTA Consolider Team,
Universitat Politècnica de València, 46022 València (Spain)

[2] Departamento de Física Aplicada-ICMUV, MALTA Consolider Team,
Universidad de Valencia, Edificio de Investigación, C/Dr. Moliner 50, 46100 Burjassot, Spain

[3] Earth Sciences Department, University College London, Gower Street, WC1E 6BT, London (UK)

[4] Geowissenschaften, Goethe-Universität, Altenhöferallee 1, 60438 Frankfurt am Main, Germany

[5] Instituto de Diseño para la Fabricación y Producción Automatizada, MALTA Consolider Team,
Universitat Politècnica de València, 46022 València, Spain

[6] CINVESTAV-Departamento de Nanociencia y Nanotecnología,
Unidad Querétaro, 76230 Querétaro, México

[7] Departamento de Física, Instituto de Materiales y Nanotecnología, MALTA Consolider Team,
Universidad de La Laguna, 38205 La Laguna, Tenerife, Spain

[†] On leave from Departamento de Física, Universidad Distrital "Francisco José de Caldas",
110311 Bogotá, Colombia

[a)] Corresponding author, email: osgohi@fis.upv.es

## Abstract

The structural properties of Thallium (III) oxide (Tl$_2$O$_3$) have been studied both experimentally and theoretically under compression at room temperature. X-ray powder diffraction measurements up to 37.7 GPa have been complemented with *ab initio* total-energy calculations. The equation of state of Tl$_2$O$_3$ has been determined and compared to related compounds. It has been found experimentally that Tl$_2$O$_3$ remains in its initial cubic bixbyite-type structure up to 22.0 GPa. At this pressure, the onset of amorphization is observed, being the sample fully amorphous at 25.2 GPa. The sample retains the amorphous state after pressure release. To understand the pressure-induced amorphization process, we have studied theoretically the possible high-pressure phases of Tl$_2$O$_3$. Although a phase transition is predicted at 5.8 GPa to the orthorhombic

Rh$_2$O$_3$-II-type structure and at 24.2 GPa to the orthorhombic α-Gd$_2$S$_3$-type structure, neither of these phases were observed experimentally, probably due to the hindrance of the pressure-driven phase transitions at room temperature. The theoretical study of the elastic behavior of the cubic bixbyite-type structure at high-pressure shows that amorphization above 22 GPa at room temperature might be caused by the mechanical instability of the cubic bixbyite-type structure which is theoretically predicted above 23.5 GPa.

I. Introduction

Thallium (III) oxide (Tl$_2$O$_3$) is a sesquioxide which occurs naturally as a rare mineral named avicennite **[1]**. Tl$_2$O$_3$ crystallizes at ambient conditions in the body-centered cubic bixbyite-type structure with space group (S.G.) Ia-3, No. 206, Z=16 **[2, 3, 4]**. Bixbyite-type Tl$_2$O$_3$ is isomorphic to the cubic structure of In$_2$O$_3$ and several rare-earth sesquioxides. Apart from the bixbyite structure, the corundum-type structure has been reported to be synthesized at high pressures and high temperatures **[5]**.

Tl$_2$O$_3$ can be applied in many technological areas **[6]**. In particular, it has been used as an electrode in high-efficiency solar cells due to its very low resistivity **[7,8]**. It has also been studied for optical communication applications because of its strong reflectance in the near infrared region (1300-1500 nm) **[9]**; however, its most promising application is in thallium oxide-based high-temperature superconductors **[10]**.

Despite its interesting technological applications, $Tl_2O_3$ is one of the less studied sesquioxides probably because of the poisonous nature of thallium. In particular, contact with moisture and acids may form poisonous soluble thallium compounds, like thallium acetate, whose contact with skin should be avoided **[11]**. Consequently, many properties of $Tl_2O_3$ are unknown. In particular, it was long thought that this compound behaves as a metallic conductor **[12-14]**; however, it has been recently shown that it is a degenerate n-type semiconductor **[15]**. This result is in good agreement with transport measurements which suggest that n-type conductivity comes from oxygen deficiency in the crystalline lattice **[13, 16-18]**. It is also in good agreement with optical measurements providing a band gap between 1.40 and 2.75 eV **[9, 12, 16]**.

Very little is known about the structural and mechanical properties of $Tl_2O_3$. The bulk moduli of both bixbyite-type and corundum-type structures are unknown. In this context, studies of $Tl_2O_3$ under compression could help in understanding its physical properties. In this work, we report an experimental and theoretical study of bixbyite-type $Tl_2O_3$ at room temperature and high-pressure (HP) by means of angle dispersive X-ray diffraction (ADXRD) measurements and *ab initio* calculations. Technical aspects of the experiments and calculations are described in Sections II and III, respectively. Results are presented and discussed in Section IV and conclusions summarized in Section V.

**II. Experimental Details**

Commercial $Tl_2O_3$ powder with 99.99% purity (Sigma-Aldrich) was crushed in a mortar with a pestle to obtain a micron-sized powder. X-ray diffraction (XRD) measurements performed at 1 atm and room temperature with a Rigaku Ultima IV diffractometer (Cu $K_\alpha$ radiation) confirmed the bixbyite-type structure of $Tl_2O_3$.

HP-ADXRD experiments at room temperature up to 37.7 GPa were carried out at beamline I15 of the Diamond Light Source using a monochromatic X-ray beam ($\lambda$ = 0.4246 Å) and a membrane-type diamond-anvil cell (DAC). $Tl_2O_3$ powder was loaded in a 150-μm diameter hole of an inconel gasket in a DAC with diamond-culet sizes of 350 μm. A 16:3:1 methanol-ethanol-water mixture was used as pressure-transmitting medium. A strip of gold was placed inside the gasket and used as the pressure sensor. Pressure was determined using the gold equation of state (EOS): $B_0$ = 167.5 GPa, and $B_0$' = 5.79, whose parameters are obtained with a third-order Birch-Murnaghan equation [19]. The X-ray beam was focused down to 30 x 30 $μm^2$ using Kickpatrick-Baez mirrors. A pinhole placed before the sample position was used as a clean-up aperture for filtering out the tail of the X-ray beam. The images were collected using a MAR345 image plate located at 350 mm from the sample. The diffraction patterns were integrated as a function of $2\theta$ using FIT2D in order to give conventional, one-dimensional diffraction profiles [20]. The indexing and refinement of the powder diffraction patterns was performed using the Unitcell [21], POWDERCELL [22] and GSAS [23, 24] program packages.

**III. Theoretical calculations**

We have performed *ab initio* total-energy calculations within the density functional theory (DFT) [25] using the plane-wave method and the pseudopotential theory with the Vienna *ab initio* simulation package (VASP) [26]. We have used the projector-augmented wave scheme (PAW) [27] implemented in this package to take into account the full nodal character of the all-electron charge density in the core region. Basis set, including plane waves up to an energy cutoff of 520 eV were used in order to achieve highly converged results and accurate description of the electronic properties.

The exchange-correlation energy was described with the generalized gradient approximation (GGA) with the PBEsol prescription [28]. A dense special k-points sampling for the Brillouin zone (BZ) integration was performed in order to obtain very well-converged energies and forces. At each selected volume, the structures were fully relaxed to their equilibrium configuration through the calculation of the forces on atoms and the stress tensor. This allows obtaining the relaxed structures at the theoretical pressures defined by the calculated stress. In the relaxed equilibrium configurations, the forces on the atoms are less than 0.006 eV/Å, and deviations of the stress tensor from a diagonal hydrostatic form are less than 1 kbar (0.1 GPa). The application of DFT-based total-energy calculations to the study of semiconductor properties under HP has been reviewed in Ref. **29**, showing that the phase stability, electronic, and dynamical properties of compounds under pressure are well described by DFT.

*Ab initio* calculations allow the study of the mechanical properties of materials. The elastic constants describe the mechanical properties of a material in the region of small deformations where the stress-strain relations are still linear. The elastic constants can be obtained by computing the macroscopic stress for a small strain with the use of the stress theorem [30]. Alternatively, the macroscopic stress can be also calculated using density functional perturbation theory (DFPT) [31]. In the present work, we perform the evaluation of the elastic constants of $Tl_2O_3$, with the use of the DFT as implemented in the VASP package [32]. The ground state and fully relaxed structures were strained in different directions according to their symmetry [32]. The total-energy variations were evaluated according to a Taylor expansion for the total energy with respect to the applied strain [33]. Due to this fact, it is important to check that the strain used in the calculations guarantees the harmonic behavior. This procedure allows us to obtain the $C_{ij}$ elastic constants in the Voigt notation. The number of independent elastic constants is reduced by crystalline symmetry [34].

IV. **Results and Discussion**

A. **X-ray diffraction and structural properties**

The crystalline structure of cubic bixbyite-type $Tl_2O_3$ (see **Fig. 1**) has two different types of six-fold-coordinated thallium atoms. Thallium located at the 8*b* Wyckoff site has slightly distorted octahedral coordination whilst thallium located at 24*d* Wyckoff site has distorted trigonal prismatic coordination. Finally, oxygen atoms occupy 48*e* Wyckoff sites. From XRD measurements carried out at 1 atm and room temperature outside the DAC we have made a Rietveld refinement of the lattice parameter and relative atomic positions of the bixbyite-type structure. The refinement *R*-values are $R_p$ = 7.7 %, and $R_{wp}$ = 10.2 %. These results (summarized in **Table I**) are in quite a good agreement with those of **Refs. 2, 3**, and **4**, and with our calculations, all of them included in **Table I** for comparison.

**Figure 2** shows ADXRD patterns of $Tl_2O_3$ up to 37.7 GPa. Diffractograms up to 22 GPa can be indexed with the cubic bixbyite-type $Tl_2O_3$ structure. The main difference between diffraction patterns up to 22 GPa is the shift of Bragg peaks to higher angles with pressure as the result of a unit-cell volume decrease. A typical peak broadening of XRD peaks [35] is detectable above 11 GPa. In this respect, before continuing the discussion of the results, we would like to comment on possible non-hydrostatic effects in our experiments. We have checked that non-hydrostatic conditions above 11 GPa do not induce a tetragonal or rhombohedral distortion of the cubic structure. As an example, the Rietveld refinement of the powder XRD pattern measured at 18.2 GPa is included in **Fig. 2**. The refined parameters were: the scale factor, phase fractions, lattice parameters, profile coefficients, *x* fractional atomic coordinate of the Tl(2) atom, the overall displacement factor, and the background. The high quality of the Rietveld refinement shows that $Tl_2O_3$ remains in the cubic phase at 18.2 GPa; i.e., just before the onset of the amorphization process, as will be explained below. We note that the broadening of XRD peaks above 11 GPa could be due to the loss of quasi-hydrostatic conditions of the pressure-transmitting medium **[36-38]** or to local distortions caused by the appearance of defects which could be precursors of the pressure-induced amorphization (PIA) that will be commented later on.

From the refinement of the diffraction patterns up to 22 GPa, we obtained the pressure dependence of the $Tl_2O_3$ lattice parameter. In the Rietveld refinement, the oxygen atomic coordinates were supposed not to vary with pressure due to its small X-ray scattering cross section in comparison to that of thallium atom. Rietveld refinements carried out on HP-ADXRD data for $Tl_2O_3$ show that the *x* fractional atomic coordinate of the Tl(2) atom up to 22 GPa was similar to that at 1 atm within experimental uncertainty. This result agrees with the weak pressure dependence of this atomic parameter obtained from our theoretical calculations (according to simulations, the *x* fractional atomic coordinate of Tl(2) in $Tl_2O_3$ varies from 0.9667 at 0 GPa to 0.9664 at 22 GPa). The pressure evolution of the unit-cell volume of $Tl_2O_3$ is plotted in **Fig. 3**. We have fitted these data with second-order Birch-Murnaghan (BM2) and third-order Birch-Murnaghan (BM3) EOSs **[39]**. Weights derived from the experimental uncertainties in both pressure and volume were assigned to each data point in both fits. The fits were carried out with the EoSFit software (v.5.2) **[40]**. All the experimental and theoretical values at zero pressure for the volume, $V_0$, bulk modulus, $B_0$, and its first-pressure derivative, $B_0$', are summarized in **Table II**. Our experimental values are in relatively good agreement with our calculated values. For the case of our experimental data, the obtained value for the weighted chi-squared, $\chi^2_w$, in the BM2 and BM3 EOS fits is 6.2 and 6.7; respectively. We note that, the refinement of the $B_0$' parameter in the BM3 EOS fit does not improve the fit of the data because the $\chi^2_w$ increases to a value of 6.7 and the standard deviation of $B_0$ increases with respect to that obtained with the BM2 EOS, thus indicating that an expansion of the EOS to third order is not required to fit the data. These results show that the second-order equation of state is an adequate representation of the volume-pressure data of $Tl_2O_3$. Furthermore, the EOS parameters for isostructural $In_2O_3$ are also included in **Table II** for comparison **[41-43]**. It can be highlighted that the experimental value for $B_0$ in $Tl_2O_3$ ($B_0$ = 156(3) GPa) is approximately 15% smaller than that obtained for $In_2O_3$ ($B_0$ = 184(10) GPa) **[43]**. In this comparison, we considered the EOS parameters obtained with a BM2 EOS with $B_0$' fixed to 4 because the $B_0$ and $B_0$' parameters are strongly correlated **[44].** The lower value of $B_0$ for cubic $Tl_2O_3$ when compared to that of $In_2O_3$ is consistent with the decrease of the bulk modulus of bixbyite-type sesquioxides when the ionic radius of the $A$ cation increases in the series $A$ = In, Tl. We note that bixbyite-type sesquioxides like $In_2O_3$ and $Tl_2O_3$ are much less compressible than sesquioxides of late group-15 elements in the Periodic Table like cubic α-$Sb_2O_3$ (S.G.: $Fd$-$3m$, No. 227, Z=16) **[45]** and monoclinic α-$Bi_2O_3$ (S.G.: $P2_1/c$, No. 14, Z=4) **[46].**

**Figure 2** shows a drastic decrease of the intensity of the Bragg reflections of cubic $Tl_2O_3$ between 18.2 and 22 GPa. In addition, at 25.2 GPa all the sharp crystalline peaks of cubic $Tl_2O_3$ disappear and two broad peaks appear at 8.75º and 11.53º (noted with asterisk marks). These two peaks remain up to 37.7 GPa, the maximum pressure achieved in our experiment, and exhibit a small shift to higher angles between 25.2 and 37.7 GPa. These results can be interpreted as an amorphization of $Tl_2O_3$ above 22.0 GPa which is already completed at 25.2 GPa, and will be discussed in the next section.

**B. Amorphization**

It is commonly accepted that PIA in crystalline solids may occur if the crystalline structure becomes mechanical or dynamically unstable at a certain pressure; i.e., if mechanical stability criteria are violated or if the phonon dispersion curves contain imaginary frequencies for phonon modes at a given pressure **[47].** PIA due to these instabilities usually occurs when the crystalline solid cannot undergo a phase transition to a HP crystalline phase at a smaller pressure than that of amorphization. The hindrance of the pressure-driven phase transition between two crystalline phases is usually due to the presence of kinetic barriers between the low- and high-pressure structures. This barrier cannot be overcome if the temperature is not high enough and consequently the transition is frustrated at low temperatures. Therefore, it is worth to investigate which could be the frustrated HP phase of $Tl_2O_3$ and at which pressure the phase transition is predicted to occur.

In order to look for candidates of HP phases of $Tl_2O_3$ we have performed total-energy calculations for $Tl_2O_3$ with the structures observed experimentally in $In_2O_3$ at different pressures and temperatures **[48-52]**. They include bixbyite-type (*Ia*-3), corundum-type (S.G.: *R-3c*, No. 167, Z=6), orthorhombic $Rh_2O_3$-II-type (S.G.: *Pbcn*, No. 60, Z=4), and orthorhombic α-$Gd_2S_3$-type (S.G.: *Pnma*, No. 62, Z=4) structures. We have also considered in our calculations the orthorhombic $Rh_2O_3$-III-type structure (S.G. *Pbca*, No. 61, Z=8), which is a high-temperature and low-pressure form of $Rh_2O_3$ [53], and two of the structures commonly found in rare-earth sesquioxides (RES) under different pressure and temperature conditions [54-58], like the monoclinic B-RES (S.G.: $C2/m$, No. 12, Z=6) and trigonal A-RES (S.G. $P-3m1$, No. 164, Z=1) structures. Finally, to complete the study, we have also considered as candidates for HP phases of $Tl_2O_3$ structures observed in transition-metal sesquioxides at different pressures and temperatures. These structures are the $Sb_2S_3$-type (S.G.: $Pnma$, No. 62, Z=4) found in $Ti_2O_3$ [59, 60], the distorted orthorhombic perovskite or $GdFeO_3$-type (S.G.: $Pnma$, No. 62, Z=4) found in $Fe_2O_3$ (hematite) [61] and the orthorhombic post-perovskite or $CaIrO_3$-type (S.G.: $Cmcm$, No. 63, Z=4) found in $Mn_2O_3$ [62].

The enthalpy difference vs. pressure diagram for the different $Tl_2O_3$ polymorphs, taking as reference the enthalpy of the bixbyite-type phase, is plotted in **Fig. 4.** Our calculations predict a phase transition from the bixbyite-type phase ($Ia$-3) to the $Rh_2O_3$-II-type phase ($Pbcn$) at 5.8 GPa, and from the $Rh_2O_3$-II-type phase to the α-$Gd_2S_3$-type phase ($Pnma$) at 24.2 GPa. This sequence of pressure-induced phase transitions for $Tl_2O_3$ is the same as for $In_2O_3$ [48, 49]. The main difference is the phase transition pressures predicted theoretically in both compounds: 5.8 GPa (7-11 GPa) and 24.2 GPa (36-40 GPa) for $Tl_2O_3$ ($In_2O_3$) [48, 49]. The fact that our HP-ADXRD measurements in $Tl_2O_3$ up to 37.7 GPa do not show evidence of the phase transitions predicted at 5.8 GPa and 24.2 GPa, but an amorphization whose onset is around 22 GPa, suggests that kinetic barriers might be present in the phase transition to the HP phases at room temperature in $Tl_2O_3$. Note that the phase transition from the bixbyite-type to the $Rh_2O_3$-II-type structure was observed in $In_2O_3$ at room temperature above 30 GPa [43] and the transition to the α–$Gd_2S_3$-type structure in $In_2O_3$ was not observed at room temperature between 1 atm and 51 GPa [49]. In fact, an amorphous halo was observed in $In_2O_3$ at 51 GPa and room temperature; thus suggesting a PIA in $In_2O_3$ at room temperature above this pressure. On the other hand, the phase transition in $In_2O_3$ from the $Rh_2O_3$-II-type structure to the α–$Gd_2S_3$-type structure was observed at HP and high temperature **[49]**. Furthermore, the phase transition pressures in $In_2O_3$ were observed experimentally close to those theoretically predicted only in HP and high temperature experiments **[48, 49]**. Those works already showed that large kinetic barriers are present in $In_2O_3$ at room temperature between the bixbyite-type, $Rh_2O_3$-II-type and α–$Gd_2S_3$-type structures; therefore, similar barriers are expected to occur for the same structures in $Tl_2O_3$. In particular, the hypothesis of the kinetic frustration of the pressure-induced phase transition from the bixbyite-type to the $Rh_2O_3$-II-type structure in $Tl_2O_3$ will be explored in detail in future simultaneous HP and high temperature experiments on $Tl_2O_3$, as it was already done for $In_2O_3$ **[48-52]**. We note that our *ab initio* calculations do not include kinetic energy barriers and therefore the theoretically predicted HP phases for bixbyite-type $Tl_2O_3$ could be found experimentally in future HP and high temperature experiments where the kinetic energy barriers can be overcome. Finally, we want to stress that $Tl_2O_3$ remains amorphous after decompression from 37.7 GPa to 1 atm, i.e. PIA in $Tl_2O_3$ at room temperature is irreversible.

**C. Elastic properties**

In order to further understand the amorphization process in $Tl_2O_3$ we have studied the mechanical stability of the cubic bixbyite-type (Ia-3) structure of $Tl_2O_3$ at HP. This structure belongs to the cubic Laue group CII with point group m-3 which has three independent second order elastic constants: $C_{11}$, $C_{12}$, and $C_{44}$. **Table III** summarizes the values of the three $C_{ij}$ in $Tl_2O_3$ at zero pressure as obtained from our *ab initio* calculations. The calculated elastic constants of bixbyite-type $In_2O_3$ taken from Ref. **[63]** are also included in **Table III** for comparison. The values of the three elastic constants of $Tl_2O_3$ are smaller than those of $In_2O_3$. This result supports the smaller zero pressure bulk modulus of $Tl_2O_3$ when compared to that of $In_2O_3$ as previously commented.

A lattice is mechanically stable at zero pressure only if the Born stability criteria are fulfilled **[64]**. In the case of cubic systems these criteria are:

$$C_{11} + 2C_{12} > 0, \; C_{11} - C_{12} > 0, \; C_{44} > 0 \qquad (1)$$

In our particular case, all the above criteria are satisfied for bixbyite-type $Tl_2O_3$ at zero pressure; therefore, cubic $Tl_2O_3$ is mechanically stable at 1 atm ($10^{-4}$ GPa), as it was expected. When a non-zero uniform stress is applied to the crystal, the above criteria to describe the stability limits of the crystal at finite strain are not adequate and the Born stability criteria must be modified. In this case, the elastic stiffness, or stress-strain, coefficients are defined as

$$B_{ijkl} = C_{ijkl} + 1/2 \, [ \, \delta_{ik} \sigma_{jl} + \delta_{jk} \sigma_{il} + \delta_{il} \sigma_{jk} + \delta_{jl} \sigma_{ik} - 2 \, \delta_{kl} \sigma_{ij} ] \qquad (2)$$

where the $C_{ijkl}$ are the elastic constants evaluated at the current stressed state, $\sigma_{ij}$ correspond to the external stresses, and $\delta_{kl}$ is the Kronecker delta **[65-67]**. In the special case of hydrostatic pressure applied to a cubic crystal, $\sigma_{11} = \sigma_{22} = \sigma_{33} = -P$, and the elastic stiffness coefficients are: $B_{11} = C_{11} - P$, $B_{12} = C_{12} + P$, and $B_{44} = C_{44} - P$, where $P$ is the hydrostatic pressure. Note that the $B_{ij}$ and $C_{ij}$ coefficients are equal at 0 GPa. When the $B_{ij}$ elastic stiffness coefficients are used, all the relations of the theory of elasticity can be applied including Born's stability conditions which are identical in both loaded and unloaded states **[66-69]**.

The bulk (*B*), and shear (*G*) moduli of cubic Tl$_2$O$_3$ can be obtained in the Voigt [70], Reuss [71] and Hill [72] approximations, labeled with subscripts *V*, *R*, and *H*, respectively, using the formulae [73]:

$$B_V = B_R = \frac{B_{11} + 2B_{12}}{3} \tag{3}$$

$$B_H = \frac{B_V + B_R}{2} \tag{4}$$

$$G_V = \frac{B_{11} - B_{12} + 3B_{44}}{5} \tag{5}$$

$$G_R = \frac{5(B_{11} - B_{12})B_{44}}{4B_{44} + 3(B_{11} - B_{12})} \tag{6}$$

$$G_H = \frac{G_V + G_R}{2} \tag{7}$$

In the Voigt (Reuss) approximation, uniform strain (stress) is assumed throughout the polycrystal [70, 71]. On the other hand, Hill has shown that the Voigt and Reuss averages are limits and suggested that the actual effective *B* and *G* elastic moduli can be approximated by the arithmetic mean of the two bounds [72]. The Young (*E*) modulus and the Poisson's ratio (ν) are given by [74, 75]:

$$E_X = \frac{9B_X G_X}{G_X + 3B_X} \tag{8}$$

$$\nu_X = \frac{1}{2}\left(\frac{3B_X - 2G_X}{3B_X + G_X}\right) \tag{9}$$

where the subscript *X* refers to the symbols *V*, *R*, and *H*. We summarize in **Table III** all the values obtained for *B*, *G*, *E*, and ν in bixbyite-type Tl$_2$O$_3$ at zero pressure in the Voigt, Reuss, and Hill approximations. Note that our calculated value for the bulk modulus in the Hill approximation ($B_H$ = 125.1 GPa) is in very good agreement with the value of $B_0$ = 125.0(4) GPa obtained from our PBEsol structural calculations via a BM3

EOS fit. This result gives us confidence about the correctness of our elastic constants calculations.

**Table III** also includes the values of the ratio between the bulk and shear modulus, $B/G$, and the Zener anisotropy factor, $A$. The $B/G$ ratio has been proposed by Pugh to predict brittle or ductile behavior of materials **[76]**. According to the Pugh criterion, a $B/G$ value above 1.75 indicates a tendency for ductility; otherwise, the material behaves in a brittle manner. In our particular case, we found a value of $B/G = 3.56$ in the Hill approximation indicating that the material should be ductile at 1 atm. The Zener anisotropy factor $A$ for our cubic cell is defined as $A=2B_{44}/(B_{11}-B_{12})$. If $A$ is equal to one, no anisotropy exists. On the other hand, the more this parameter differs from one, the more elastically anisotropic is the crystalline structure. In cubic $Tl_2O_3$, the $A$ value (0.84) is slightly different from 1 and evidence a small elastic anisotropy of our cubic cell at 1 atm.

**Figures 5(a)** and **5(b)** show the pressure dependence of the three calculated $C_{ij}$ elastic constants and the three $B_{ij}$ elastic stiffness coefficients of bixbyite-type $Tl_2O_3$, respectively. It can be seen that $B_{11}$ and $B_{12}$ increase monotonically as pressure increases, while $B_{44}$ decreases monotonically as pressure increases and at 23.5 GPa crosses the 0 GPa horizontal line. This fact is related with the mechanical instability of bixbyite-type $Tl_2O_3$ and will be discussed in the next paragraphs.

The knowledge of the behavior of the three elastic stiffness coefficients with pressure allows us to study the mechanical stability of bixbyite-type $Tl_2O_3$ as pressure increases. The new conditions for elastic stability at a given pressure $P$, known as the generalized stability criteria, are obtained by replacing in **Eq. (1)** the $C_{ij}$ elastic constants by the $B_{ij}$ elastic stiffness coefficients, and are given by **[77]**:

$$M_1 = B_{11} + 2B_{12} > 0 \qquad (10)$$

$$M_2 = B_{11} - B_{12} > 0 \tag{11}$$

$$M_3 = B_{44} > 0. \tag{12}$$

where $B_{11}$, $B_{12}$, and $B_{14}$ are the elastic stiffness coefficients at the considered pressure. These generalized stability criteria are plotted in **Fig. 6**. It is found that **Eq. (12)**, related to a pure shear instability, is violated at 23.5 GPa while **Eq. (11)**, called the Born instability **[77]**, is violated at 26.0 GPa. Therefore, our theoretical study of the mechanical stability of $Tl_2O_3$ at HP suggests that the bixbyite-type phase becomes mechanically unstable beyond 23.5 GPa. This pressure is slightly above but very close to the pressure at which the onset of the amorphization process takes place experimentally. Consequently, this result suggests that shear instability could be involved in the amorphization process of $Tl_2O_3$ at room temperature. Note that our calculations are performed for a perfect material, whereas our powder samples are very defective and contain a high concentration of O vacancies that make $Tl_2O_3$ a degenerate n-type semiconductor. Therefore, we expect that the amorphization in our sample takes place at a lower pressure than that theoretically predicted since defects are known to induce amorphization and decrease the pressure at which the amorphization process begins in a number of materials **[78, 79]**.

We have performed the study of the dynamical stability in $Tl_2O_3$ in order to complement the study of the mechanical stability of $Tl_2O_3$ and verify that PIA in $Tl_2O_3$ is caused by the mechanical instability of the cubic phase. To check the dynamical stability of the cubic phase, we have carried out *ab initio* calculations of the phonon dispersion relations in bixbyite-type $Tl_2O_3$. We have found that the cubic phase is dynamically stable up to 32 GPa and that phonons with imaginary frequencies appear above this pressure. This result thus indicates that bixbyite-type $Tl_2O_3$ becomes dynamically unstable above 32 GPa **[80]**. Since this pressure is higher than the pressure at which bixbyite-type $Tl_2O_3$ becomes mechanically unstable, we conclude that PIA of $Tl_2O_3$ observed at room temperature at 22 GPa might be caused by the mechanical instability of the cubic lattice at pressures above 22 GPa.

Finally, for completeness we have plotted the pressure dependence of the elastic moduli ($B_H$, $G_H$ and $E_H$), $v_H$ Poisson's ratio, $B_H/G_H$ ratio, and $A$ Zener anisotropy factor in **Fig. 7**. It is found that $B_H$ increases with pressure and reaches the value of 216.0 GPa at 23 GPa. On the other hand, $G_H$ and $E_H$ decrease with pressure approaching a value of 0 GPa near 23.5 GPa, pressure at which the mechanical instability is predicted to occur. We note that the fact that the shear modulus decreases with pressure is compatible with the fact that the equation that first is violated (**Eq. 12**) is the one related with the pure shear instability because of the decreasing of $B_{44}$ with pressure. The Poisson's ratio, $v_H$, increases with pressure and reaches a value of 0.49 at 23 GPa. The $B_H/G_H$ ratio increases with pressure, grows exponentially above 19 GPa, and reaches a value of 94.4 at 23 GPa. The increase of the $B_H/G_H$ ratio with pressure indicates that the ductility of $Tl_2O_3$ is enhanced under compression. In the case of the Zener anisotropy factor, $A$, it is found that it increases with pressure reaching a maximum value of $A = 0.96$ at about 11 GPa and afterward decreases quickly above 20 GPa indicating a strong increase of the elastic anisotropy above that pressure.

## V. Concluding Remarks

We have studied both experimentally and theoretically the structural properties of $Tl_2O_3$ under compression at room temperature. The equation of state of $Tl_2O_3$ has been determined and its bulk modulus has been found to be smaller than that of isostructural $In_2O_3$. $Tl_2O_3$ starts to amorphize above 22 GPa and retains the amorphous structure at 1 atm when decreasing pressure from 37.7 GPa. The theoretically predicted transitions to the $Rh_2O_3$-II-type structure, near 6 GPa, and to the $\alpha$-$Gd_2S_3$-type structure, near 24 GPa, are not observed experimentally, probably, due to the kinetic hindrance of the phase transitions at room temperature.

To understand the pressure-induced amorphization process of $Tl_2O_3$, we have studied theoretically both the mechanical and dynamical stability of the cubic phase at high pressures. In this respect, the mechanical properties of bixbyite-type $Tl_2O_3$ at high pressures have been commented. Our calculations show that amorphization might be caused by the mechanical instability of the bixbyite-type structure predicted above 23.5 GPa since the cubic phase is dynamically stable up to 32 GPa.

**Acknowledgments**

This study was supported by the Spanish government MEC under Grants No: MAT2010-21270-C04-01/03/04, MAT2013-46649-C4-1/2/3-P and CTQ2009-14596-C02-01, by the Comunidad de Madrid and European Social Fund (S2009/PPQ-1551 4161893), by MALTA Consolider Ingenio 2010 project (CSD2007-00045), and by Generalitat Valenciana (GVA-ACOMP-2013-1012 and GVA-ACOMP-2014-243). We acknowledge Diamond Light Source for time on beamline I15 under proposal EE6517 and I15 beamline scientist for technical support. A.M. and P.R-H. acknowledge computing time provided by Red Española de Supercomputación (RES) and MALTA-Cluster. B.G.-D. and J.A.S. acknowledge financial support through the FPI program and Juan de la Cierva fellowship. J.R.-F. acknowledges the Alexander von Humboldt Foundation for a postdoctoral fellowship.

**References**

[1] Kh. N. Karpova, E. A. Kon'kova, E. D. Larkin, and V. F. Savel'ev, Doklady Akad. Nauk. Uzbekistan S.S.R. **2**, 23 (1958).

[2] P. Papamantellos, Z. Kristallogr. **126**, 143 (1968).

[3] H. H. Otto, R. Baltrusch, and H.-J. Brandt, Physica C **215**, 205 (1993).

[4] P. Berastegui, S. Eriksson, S. Hull, F.J. García-García, and J. Eriksen, Solid State Sciences **6**, 433 (2004).

[5] C.T. Prewitt, R.D. Shannon, D.B. Rogers, and A.W. Sleight, Inorg. Chem. **8**, 1985 (1969).

[6] C. R. Patra, A. Gedanken, New J. Chem. **28**, 1060 (2004).

[7] J.A. Switzer, J. Electrochem. Soc. **133**, 722 (1986).

[8] R.J. Phillips, M.J. Shane and J.A. Switzer, J. Mat. Res. **4**, 923 (1989).

[9] R.A. Van Leeuwen, C.J. Hung, D.R. Kammler, and J.A. Switzer, J. Phys. Chem. **99**, 15247 (1995).

[10] R.N. Bhattacharya, Thallium-oxide superconductors, in: R.N. Bhattacharya and M. Parans Paranthaman (Eds.), *High Temperature Superconductors*, pp. 129-151 (Wiley-VCH, Weinheim, 2010).

[11] C.D. Weaver, D. Harden, S.I Dworetzky, B. Robertson, and R.J Knox, J. Biomol. Screen **9**, 671 (2004).

[12] H. P. Geserich, phys. stat. sol. **25**, 741 (1968).

[13] A. Goto, H. Yasuoka, A. Hayashi, and Y. Ueda, J. Phys. Soc. Jpn. **61**, 1178 (1992).

[14] P. A. Glans, T. Learmonth, K.E. Smith, et al., Phys. Rev. B **71**, 235109 (2005).

[15] A. B. Kehoe, D. O. Scanlon, G. W. Watson, Phys. Rev. B **83**, 233202 (2011).

[16] V.N. Shukla and G.P. Wirtz, J. Am. Ceram. Soc. **60**, 253 (1977).

[17] V.N. Shukla and G.P. Wirtz, J. Am. Ceram. Soc. **60**, 259 (1977).

[18] G.P. Wirtz, C.J. Yu, and R.W. Doser, J. Am. Ceram. Soc. **64**, 269 (1981).

[19] M. Yokoo, N. Kawai, K. G. Nakamura, K. I. Kondo, Y. Tange, and T. Tsuchiya, Phys. Rev. B **80**, 104114 (2009).

[20] A. P. Hammersley, S. O. Svensson, M. Hanfland, A. N. Fitch, and D. Häusermann, High Pressure Research, **14,** 235 (1996).

[21] T. J. B. Holland, and S. A. T. Redfern, Mineralogical Magazine, **61**, 65 (1997).

[22] W. Kraus and G. Nolze, J. Appl. Crystallogr. **29,** 301 (1996).

[23] A. C. Larson and R. B. von Dreele, LANL Report 86-748, (2004).

[24] B. H. Toby, J. Appl. Cryst. **34**, 210 (2001).

[25] P. Hohenberg and W. Kohn, Phys. Rev. **136**, B864 (1964).

[26] G. Kresse and J. Hafner, Phys. Rev. B **47**, 558 (1993); G. Kresse and J. Hafner, Phys. Rev. B **49**, 14251 (1994); G. Kresse and J. Furthmüller, Comput. Mat. Sci. **6**, 15 (1996); G. Kresse and J. Furthmüller, Phys. Rev. B **54**, 11169 (1996).

[27] P. E. Blöchl, Phys. Rev. B **50**, 17953 (1994); G. Kresse and D. Joubert, Phys. Rev. B **59**, 1758 (1999).

[28] J. P. Perdew, A. Ruzsinszky, G. I. Csonka, O. A. Vydrov, G. E. Scuseria, L. A. Constantin, X. Zhou, and K. Burke, Phys. Rev. Lett. **100**, 136406 (2008).

[29] A. Mujica, A. Rubio, A. Muñoz, and R. J. Needs, Rev. Mod. Phys. **79**, 863 (2003).

[30] N. Chetty, A. Muñoz, and R. M. Martin, Phys. Rev. B **40**, 11934 (1989).

[31] S. Baroni, S. de Gironcoli, A. Dal Corso, and P. Giannozzi, Rev. Mod. Phys. **73**, 515 (2001).

[32] Y. Le Page and P. Saxe, Phys. Rev. B **65**, 104104 (2002).

[33] O. Beckstein, J. E. Klepeis, G. L. W. Hart, and O. Pankratov, Phys. Rev. B **63**, 134112 (2001).

[34] J. F. Nye, *Physical properties of crystals. Their representation by tensor and matrices.* (Oxford University Press, 1957).

[35] O. Gomis, J. A. Sans, R. Lacomba-Perales, D. Errandonea, Y. Meng, J. C. Chervin, and A. Polian, Phys. Rev. B **86**, 054121 (2012).

[36] D. He and T. S. Duffy, Phys. Rev. B **73**, 134106 (2006).

[37] D. Errandonea, R. Boehler, S. Japel, M. Mezouar, and L. R. Benedetti, Phys. Rev. B **73**, 092106 (2006).

[38] S. Klotz, J. C. Chervin, P. Munsch, and G. Le Marchand, J. Phys. D: Appl. Phys. **42**, 075413 (2009)

[39] F. Birch, J. Geophys. Res. **83** (1978) 1257.

[40] R. J. Angel, Equations of state, in: R. M. Hazen and R. T. Downs (Eds.), High-temperature and high-pressure crystal chemistry. *MSA Reviews in Mineralogy and Geochemistry*, vol. 41, pp. 35–60 (Mineralogical Society of America, 2000).

[41] D. Liu, W. W. Lei, B. Zou, S. D. Yu, J. Hao, K. Wang, B. B. Liu, Q. L. Cui, and G. T. Zou, J. Appl. Phys. **104**, 083506 (2008).

[42] J. Qi, J. F. Liu, Y. He, W. Chen, and C. Wang, J. Appl. Phys. **109**, 063520 (2011).

[43] B. García-Domene, J. A. Sans, O. Gomis, F. J. Manjón, H. M. Ortiz, D. Errandonea, D. Santamaría-Pérez, D. Martínez-García, R. Vilaplana, A. L. J. Pereira, A. Morales-García, P. Rodríguez-Hernández, A. Muñoz, C. Popescu, and A. Segura, Journal of Physical Chemistry C **118,** 20545 (2014).

[44] R. J. Angel, J. L. Mosenfelder, and C. S. J. Shaw, Phys. Earth Planet. Inter. 124, 71, (2001).

[45] A. L. J. Pereira, L. Gracia, D. Santamaría-Pérez, R. Vilaplana, F. J. Manjón, D. Errandonea, M. Nalin, and A. Beltrán, Phys. Rev. B **85**, 174108 (2012).

[46] A. L. J. Pereria, D. Errandonea, A. Beltrán, L. Gracia, O. Gomis, J. A. Sans, B. García-Domene, A. Miquel-Veyrat, F. J. Manjón, A. Muñoz, and C. Popescu, J. Phys.: Condens. Matter **25**, 475402 (2013).

[47] N. Choudhury, and S. L. Chaplot, Phys. Rev. B **73**, 094304 (2006).

[48] H. Yusa, T. Tsuchiya, N. Sata, and Y. Ohishi, Physical Review B **77** 064107 (2008).

[49] H. Yusa, T. Tsuchiya, J. Tsuchiya, N. Sata, and Y. Ohishi, Physical Review B, **78** 092107 (2008).

[50] A. Gurlo, D. Dzivenko, P. Kroll, and R. Riedel, Phys. stat. sol (RRL) **2**, 269 (2008).

[51] M. F. Bekheet, M. R. Schwarz, S. Lauterbach, H. J. Kleebe, P. Kroll, A. Stewart, U. Kolb, R. Riedel, and A. Gurlo, High pressure research **33**, 697 (2013).

[52] M. F. Bekheet, M. R. Schwarz, S. Lauterbach, H. J. Kleebe, P. Kroll, R. Riedel, and A. Gurlo, Angew. Chem. Int. Ed. **52**, 6531 (2013).

[53] J. W. M. Biesterbos, J. Hornstra, Journal of the Less-Common Metals **30**, 121 (1973).

[54] L. Wang, Y. Pan, Y. Ding, W. Yang, W. L. Mao, S. V. Sinogeikin, Y. Meng, G. Shen, and H.-K. Mao, Appl. Phys. Lett. **94**, 061921 (2009).

[55] E. Husson, C. Proust, P. Guillet, and J. P. Itié, Materials Research Bulletin **34,** 2085 (1999).

[56] C. Meyer, J. P. Sanchez, J. Thomasson, and J. P. Itié, Physical Review B **51**, 12187 (1995).

[57] Q. Guo, Y. Zhao, C. Jiang, W. L. Mao, Z. Wang, J. Zhang, and Y. Wang, Inorganic Chemistry **46**, 6164 (2007).

[58] Q. Guo, Y. Zhao, C. Jiang, W. L. Mao, and Z. Wang, Solid State Communications **145**, 250 (2008).

[59] D. Nishio-Hamane, M. Katagiri, K. Niwa, A. Sano-Furukawa, T. Okada, and T. Yagi, High pressure research **29**, 379 (2009).

[60] S. V. Ovsyannikov, X. Wu, V. V. Shchennikov, A. E. Karkin, N. Dubrovinskaia, G. Garbarino, and L. Dubrovinsky, J. Phys.: Condens. Matter **22**, 375402 (2010).

[61] S. Ono, K. Funakoshi, Y. Ohishi, and E. Takahashi, J. Phys.: Condens. Matter **17**, 269 (2005).

[62] J. Santillán, S. H. Shim, G. Shen, and V. B. Prakapenka, Geophys. Res. Lett. **33**, L15307 (2006).

[63] H. Yao, L. Ouyang, and W.-Y. Ching, J. Am. Ceram. Soc. **90**, 3194 (2007).

[64] M. Born, Proc. Cambridge Philos. Soc. **36**, 160 (1940).

[65] D. C. Wallace, Thermoelastic Theory of Stressed Crystals and Higher-Order Elastic Constants, in: F.S. Henry Ehrenreich, D. Turnbull and F. Seitz (Eds.), *Solid State Physics*, vol. 25, pp. 301–404 (Academic Press, 1970).

[66] J. Wang, S. Yip, S. R. Phillpot, and D. Wolf, Phys. Rev. Lett. **71**, 4182 (1993).

[67] J. Wang, J. Li, S. Yip, S. Phillpot, and D. Wolf, Phys. Rev. B **52**, 12627 (1995).

[68] B. B. Karki, L. Stixrude, and R. M. Wentzcovitch, Reviews of Geophysics **39**, 507 (2001).

[69] O. M. Krasil'nikov, M. P. Belov, A. V. Lugovskoy, I. Yu. Mosyagin, and Yu. Kh. Vekilov, Computational Materials Science **81**, 313 (2014).

[70] W. Voigt, *Lehrbuch der Kristallphysik* (Teubner, Leipzig, 1928).

[71] A. Reuss, and Z. Angew, Math. Mech. **9**, 49 (1929).

[72] R. Hill, Proc. Phys. Soc. London, A **65**, 349 (1952).

[73] Z.-J. Wu, E.-J. Zhao, H.-P. Xiang, X.-F. Hao, X.-J. Liu, and J. Meng, Phys. Rev. B **76**, 054115 (2007).

[74] R. Caracas, and T. B. Ballaran, Physics of the Earth and Planetary Interiors, **181**, 21 (2010).

[75] Q.J. Liu, Z. T. Liu, and L. P. Feng, Commun. Theor. Phys. **56**, 779 (2011).

[76] S. F. Pugh, Philos. Mag. **45**, 823 (1954).

[77] G. Grimvall, B. Magyari-Köpe, V. Ozolinš, and K. A. Persson, Rev. Mod. Phys. **84**, 945 (2012).

[78] S.M Sharma, and S.K. Sikka, Prog. Mat. Sci. **40**, 1 (1996).

[79] P. Richet and P. Gillet, Eur. J. Miner. **9**, 907 (1997).

[80] See supplementary material at [URL will be inserted by AIP] for *ab initio* phonon dispersion calculations at 0 and 34 GPa.

**Table I:** Structural parameters of bixyite-type $Tl_2O_3$ at 1 atm.

|  | X-ray diffraction[a] | *Ab initio* PBEsol[b] | Neutron diffraction[c] | X-ray diffraction[d] | Neutron diffraction[e] |
|---|---|---|---|---|---|
| $a$ (Å) | 10.5390(4) | 10.6074 | 10.543 | 10.5344(3) | 10.5363 |
| Tl(1) site: 8$b$ | $x = 0.25$ $y = 0.25$ $z = 0.25$ | $x = 0.25$ $y = 0.25$ $z = 0.25$ | $x = 0.25$ $y = 0.25$ $z = 0.25$ | $x = 0.25$ $y = 0.25$ $z = 0.25$ | $x = 0.25$ $y = 0.25$ $z = 0.25$ |
| Tl(2) site: 24$d$ | $x = 0.969(1)$ $y = 0$ $z = 0.25$ | $x = 0.9667$ $y = 0$ $z = 0.25$ | $x = 0.971(4)$ $y = 0$ $z = 0.25$ | $x = 0.96815(22)$ $y = 0$ $z = 0.25$ | $x = 0.9657(8)$ $y = 0$ $z = 0.25$ |
| O site: 48$e$ | $x = 0.388(5)$ $y = 0.394(3)$ $z = 0.148(3)$ | $x = 0.3829$ $y = 0.3885$ $z = 0.1540$ | $x = 0.397(5)$ $y = 0.377(6)$ $z = 0.157(5)$ | $x = 0.3824(17)$ $y = 0.3905(15)$ $z = 0.1542(18)$ | $x = 0.3897(10)$ $y = 0.3982(11)$ $z = 0.1431(12)$ |

[a] Our XRD measurements.

[b] Our calculations.

[c] Ref **2**.

[d] Ref **3.**

[e] Ref **4**.

**Table II:** Experimental (Exp.) and theoretical (Th.) EOS parameters for cubic bixbyite-type $Tl_2O_3$ at zero pressure. Last column indicates the EOS type used (BM2 = Birch-Murnaghan of $2^{nd}$ order, BM3 = Birch-Murnaghan of $3^{rd}$ order). Results for isostructural $In_2O_3$ are included for comparison.

| Compound | $V_0$ (Å$^3$) | $B_0$ (GPa) | $B_0'$ | Reference | EOS type |
|---|---|---|---|---|---|
| $Tl_2O_3$ (Exp.) | 1170.6(1) | 147(13) | 5(2) | This work | BM3 |
| $Tl_2O_3$ (Exp.) | 1170.6(1) | 156(3) | 4 (fixed) | This work | BM2 |
| $Tl_2O_3$ (Th.) | 1193.2(1) | 125.0(4) | 4.97(4) | This work | BM3 |
| $Tl_2O_3$ (Th.) | 1191.5(5) | 134.2(7) | 4 (fixed) | This work | BM2 |
| $In_2O_3$ (Exp.) | 1038(2) | 194(3) | 4.75 (fixed) | [41] | BM3 |
| $In_2O_3$ (Exp.) | 1035.4(2) | 178.9(9) | 5.15 | [42] | BM3 |
| $In_2O_3$ (Exp.) | 1028(2) | 184(10) | 4 (fixed) | [43] | BM2 |

**Table III.** Calculated $C_{ij}$ elastic constants and elastic moduli $B$, $G$, $E$ (in GPa) and the Poisson's ratio, ν, for $Tl_2O_3$ at zero pressure. Elastic moduli and Possion's ratio are given in the Voigt, Reuss and Hill approximations, labeled respectively with subscripts *V, R,* and *H*. The *B/G* ratio and the Zener anisotropy factor, *A*, are also given. Calculated data at zero pressure taken from **Ref. 63** for $In_2O_3$ are also added for comparison.

|  | $Tl_2O_3$[a] | $In_2O_3$[b] |
|---|---|---|
| $C_{11}$ | 177.0 | 234.3 |
| $C_{12}$ | 99.2 | 107.2 |
| $C_{44}$ | 32.8 | 62.7 |
| $B_V = B_R = B_H$ | 125.1 | 149.6 |
| $G_V$, $G_R$, $G_H$ | 35.3, 35.0, 35.1 | 63.0[c] |
| $E_V$, $E_R$, $E_H$ | 96.7, 96.1, 96.4 | 165.8[c] |
| $\nu_V$, $\nu_R$, $\nu_H$ | 0.37, 0.37, 0.37 | 0.32[c] |
| $B_V/G_V$, $B_R/G_R$, $B_H/G_H$ | 3.55, 3.57, 3.56 | 2.37[c] |
| *A* | 0.84 | 0.99 |

[a] Our calculations with GGA-PBEsol prescription.
[b] Calculated with the GGA approximation.
[c] Results calculated in the Hill approximation from reported elastic constants.

**Figure captions**

**Figure 1. (Color online)** Schematic representation of the crystalline structure of cubic bixbyite-type $Tl_2O_3$. The unit-cell and atomic bonds are shown. Oxygen corresponds to small (red) atoms while Tl(1) located at 8$b$ and Tl(2) located at 24$d$ correspond to light blue and dark blue atoms, respectively.

**Figure 2.** Room temperature XRD patterns of $Tl_2O_3$ at selected pressures. The background has not been subtracted. The diffractogram measured at 18.2 GPa is shown as empty circles. The calculated and difference XRD patterns at 18.2 GPa obtained from a Rietveld refinement are plotted with solid lines. The residuals at 18.2 GPa are $R_p$ = 2.3 % and $R_{wp}$ = 3.0%. Bragg reflections from $Tl_2O_3$ and gold are indicated with vertical ticks at 2.1 and 18.2 GPa. Gold reflections are marked with plus (+) symbols. The XRD pattern at 1 atm after releasing pressure is shown at the top.

**Figure 3: (Color online)** Evolution of the unit-cell volume with pressure. Symbols refer to experimental data. Error bars are smaller than symbol size. Red dashed line and blue dotted line represent the fit of experimental data with a BM2 and BM3 EOS; respectively. Theoretical results are plotted with solid line.

**Figure 4. (Color online)** Theoretical calculation of enthalpy difference vs. pressure for $Tl_2O_3$ polymorphs. Enthalpy of bixbyite-type phase is taken as the reference. Enthalpy is written per two formula units for all structures for the sake of comparison.

**Figure 5. (Color online)** Pressure dependence of the theoretical (a) $C_{ij}$ elastic constants and (b) $B_{ij}$ elastic stiffness coefficients of bixbyite-type $Tl_2O_3$. Solid lines connecting the calculated data points are shown as a guide to the eyes.

**Figure 6. (Color online)** $M_1$'= $M_1$/10, $M_2$, and $M_3$ stability criteria for bixbyite-type $Tl_2O_3$ as a function of pressure. The pressure for the onset of the amorphization process, $P_{am}$, in our experiments is indicated.

**Figure 7.** Pressure dependence of (a) $B_H$, (b) $G_H$, (c) $E_H$, (d) $v_H$, (e) $B_H/G_H$, and (f) $A$. Solid lines connecting the calculated data points are shown as a guide to the eyes. Results are shown in the Hill approximation.

**Figure 1**

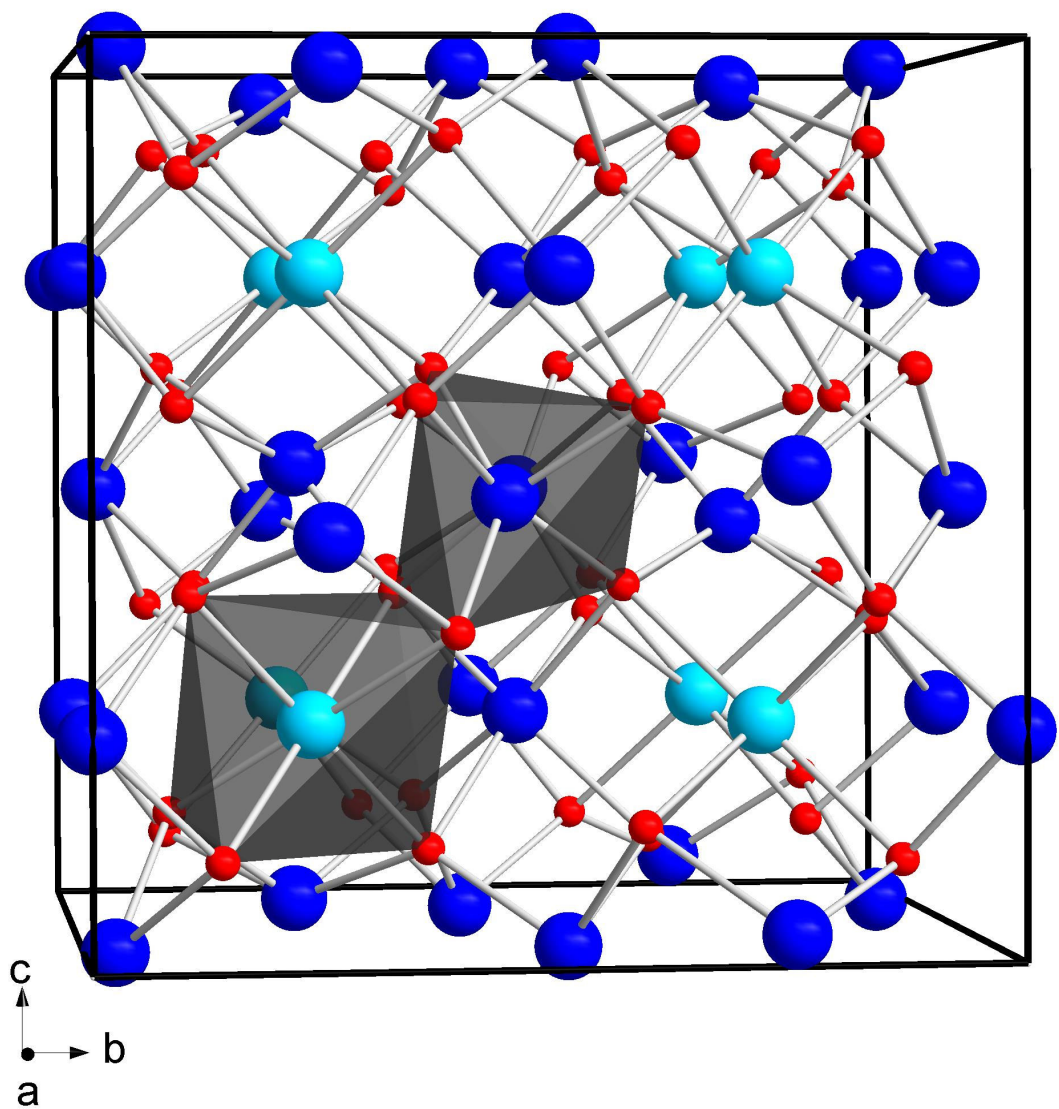

**Figure 2**

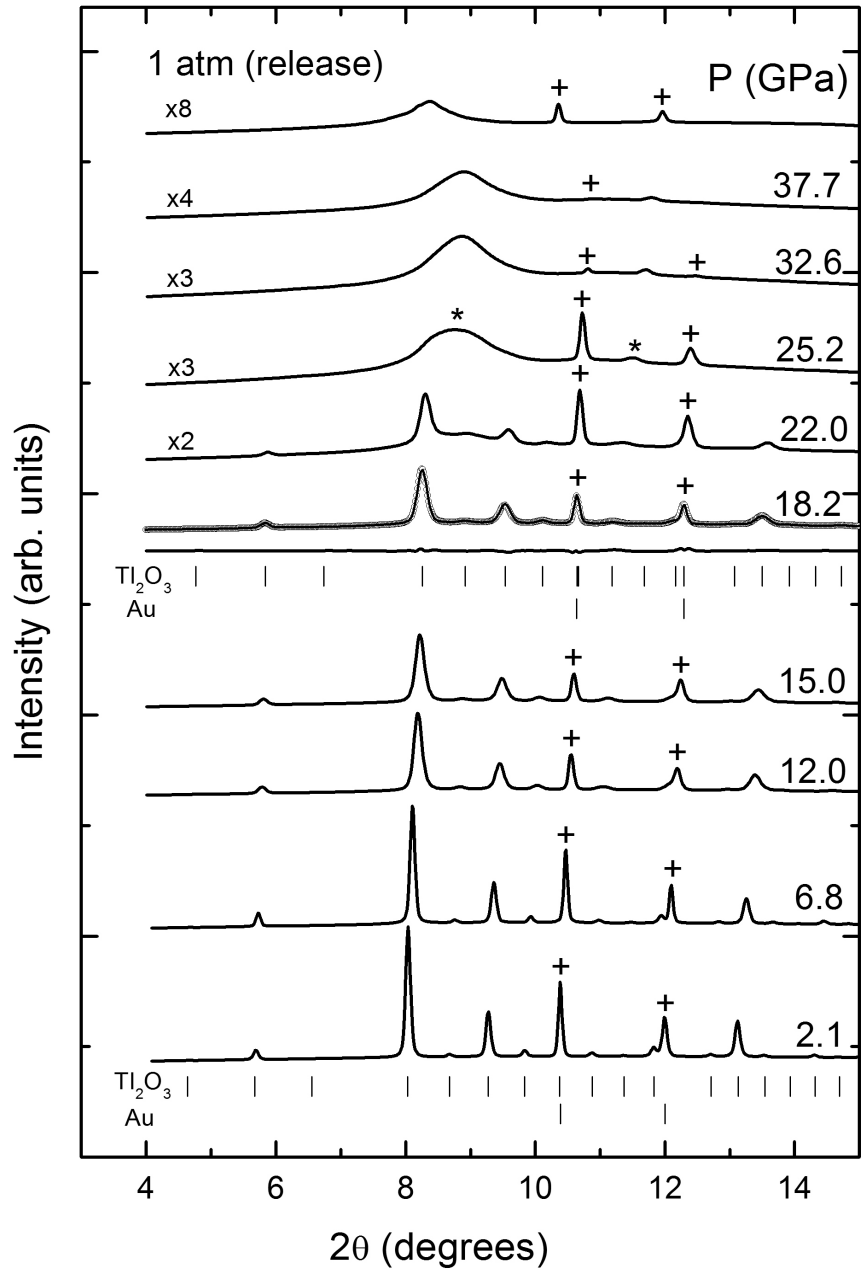

**Figure 3**

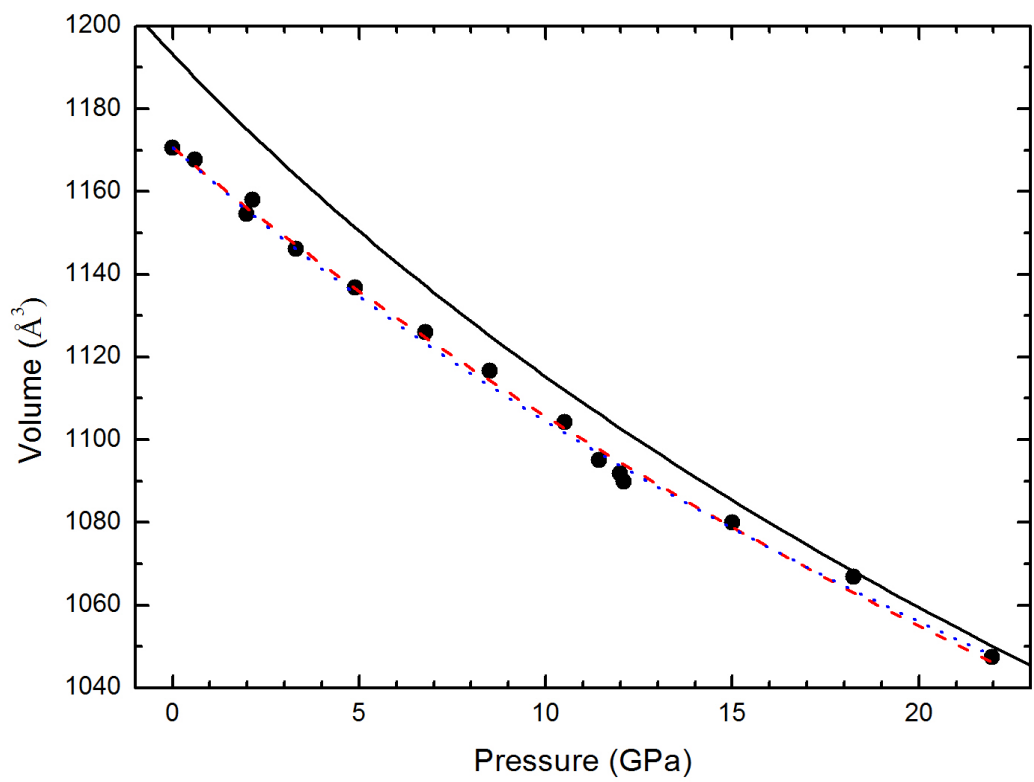

**Figure 4**

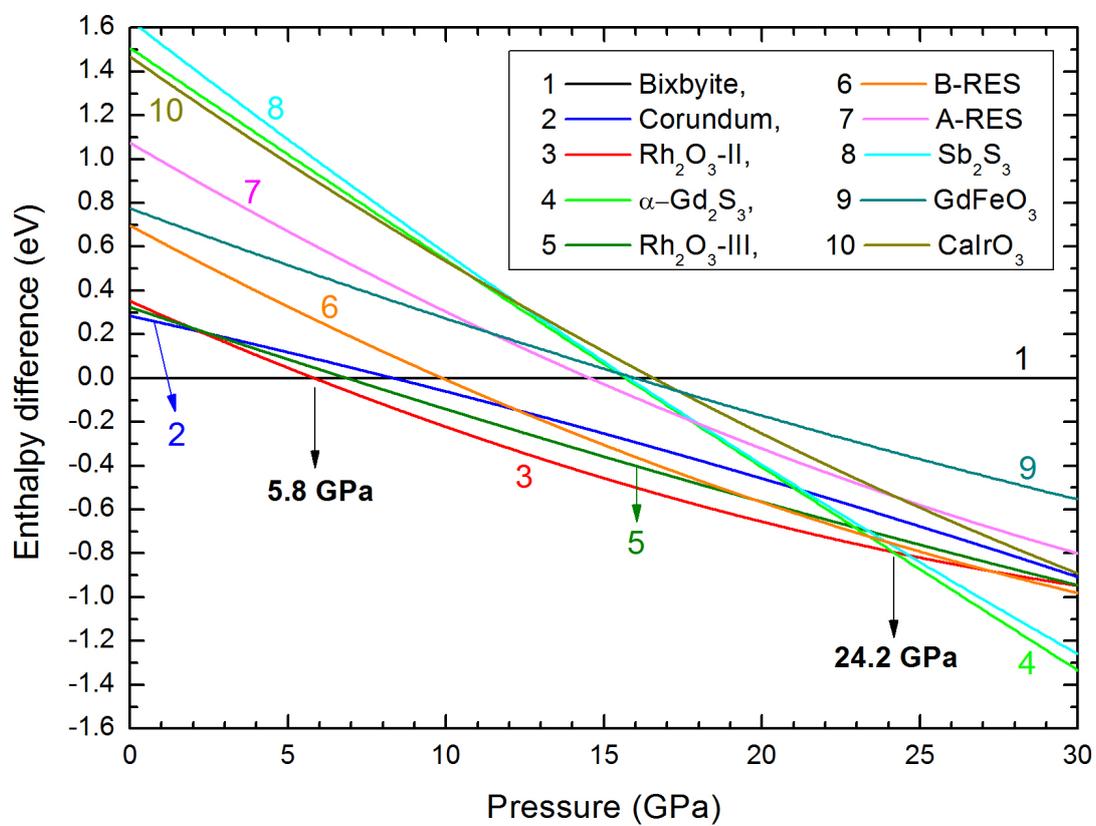

**Figure 5**

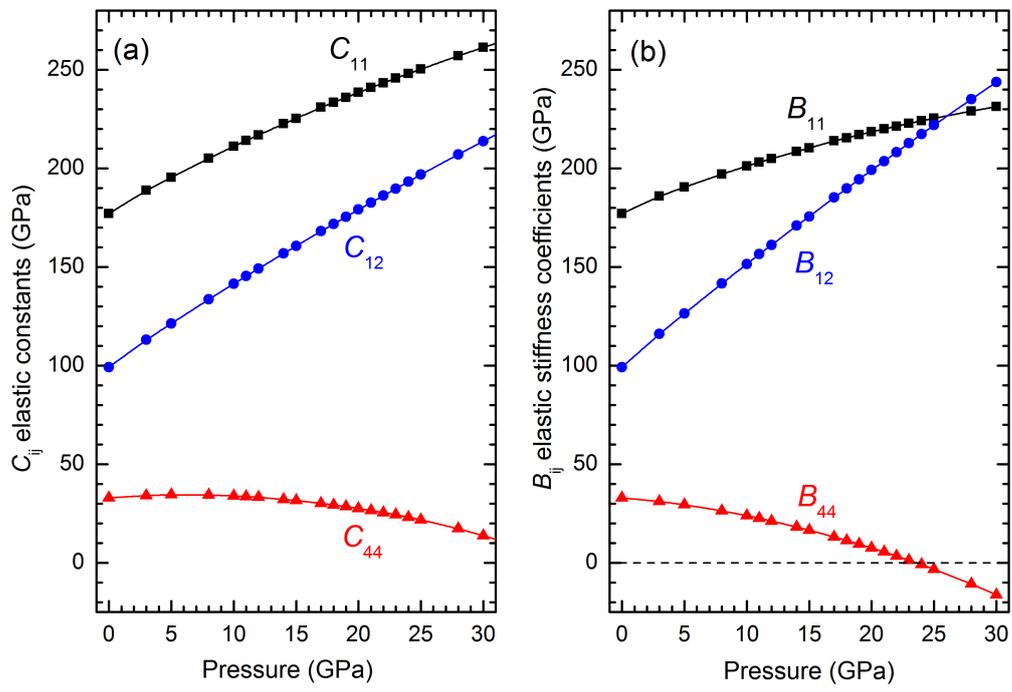

**Figure 6**

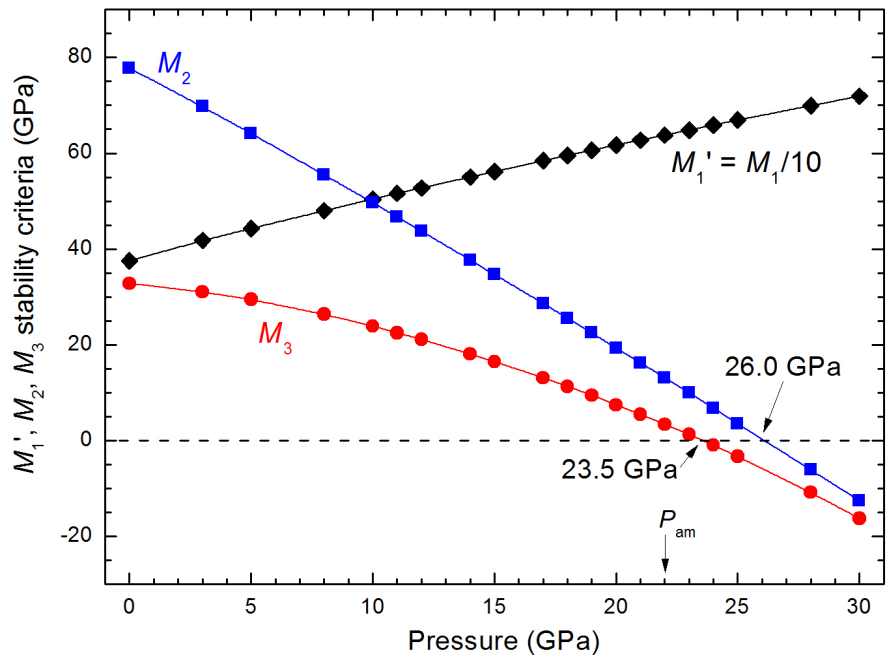

**Figure 7**

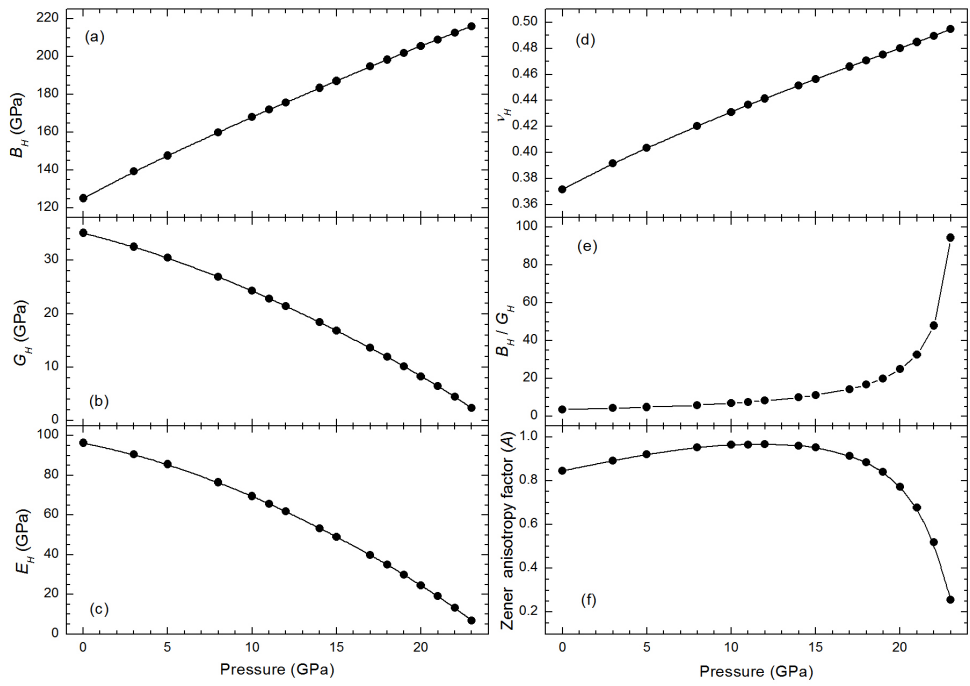

# High-pressure structural and elastic properties of Tl$_2$O$_3$

O. Gomis[1,a)] D. Santamaría-Pérez[2,3], J. Ruiz-Fuertes[2,4], J. A. Sans[5], R. Vilaplana[1], H. M. Ortiz[5,6†], B. García-Domene[2], F. J. Manjón[5], D. Errandonea[2], P. Rodríguez-Hernández[7], A. Muñoz[7], and M. Mollar[5]

[1] Centro de Tecnologías Físicas, MALTA Consolider Team,
Universitat Politècnica de València, 46022 València (Spain)

[2] Departamento de Física Aplicada-ICMUV, MALTA Consolider Team,
Universidad de Valencia, Edificio de Investigación, C/Dr. Moliner 50, 46100 Burjassot, Spain

[3] Earth Sciences Department, University College London, Gower Street, WC1E 6BT, London (UK)

[4] Geowissenschaften, Goethe-Universität, Altenhöferallee 1, 60438 Frankfurt am Main, Germany

[5] Instituto de Diseño para la Fabricación y Producción Automatizada, MALTA Consolider Team,
Universitat Politècnica de València, 46022 València, Spain

[6] CINVESTAV-Departamento de Nanociencia y Nanotecnología,
Unidad Querétaro, 76230 Querétaro, México

[7] Departamento de Física, Instituto de Materiales y Nanotecnología, MALTA Consolider Team,
Universidad de La Laguna, 38205 La Laguna, Tenerife, Spain

[†] On leave from Departamento de Física, Universidad Distrital "Francisco José de Caldas",
110311 Bogotá, Colombia

[a)] Corresponding author, email: osgohi@fis.upv.es

## Supplementary Material

First principles lattice-dynamics calculations based on DFT were performed with the PHONON code both at the zone centre (Γ point) and along high symmetry directions of the Brillouin zone (BZ) **[1]**. Highly converged results on forces are required for the calculation of the dynamical matrix using the harmonic approximation with the direct force constant approach **[1, 2]**. The construction of the dynamical matrix at the Γ point of the BZ involves separate calculations of the forces in which a fixed small displacement from the equilibrium configuration of the atoms within the relaxed cell is considered. The number of such independent displacements for the different structures is reduced due to the crystal symmetry. Diagonalization of the dynamical matrix provides the frequencies of the normal modes. Moreover, these calculations allow identifying the symmetry and eigenvectors of the phonon modes in each structure at the Γ point. In order to obtain the phonon dispersion curves we performed supercell calculations of size 2x2x2 (this means 640 atoms in the supercell) in order to obtain a good description of the phonon branches. **Figure S1** shows the phonon dispersion curves of bixbyite-type Tl$_2$O$_3$ at 0 and 34 GPa. It can be observed that at 34 GPa there are phonons with imaginary frequencies around the H point and between the Γ and P points of the BZ.

[1] K. Parlinsky, Computer Code PHONON. http:// wolf.ifj.edu.pl/phonon.

[2] K. Parlinsky, Z. Q. Li, and Y. Kawazoe, Phys. Rev. Lett. **78** 4063 (1997).

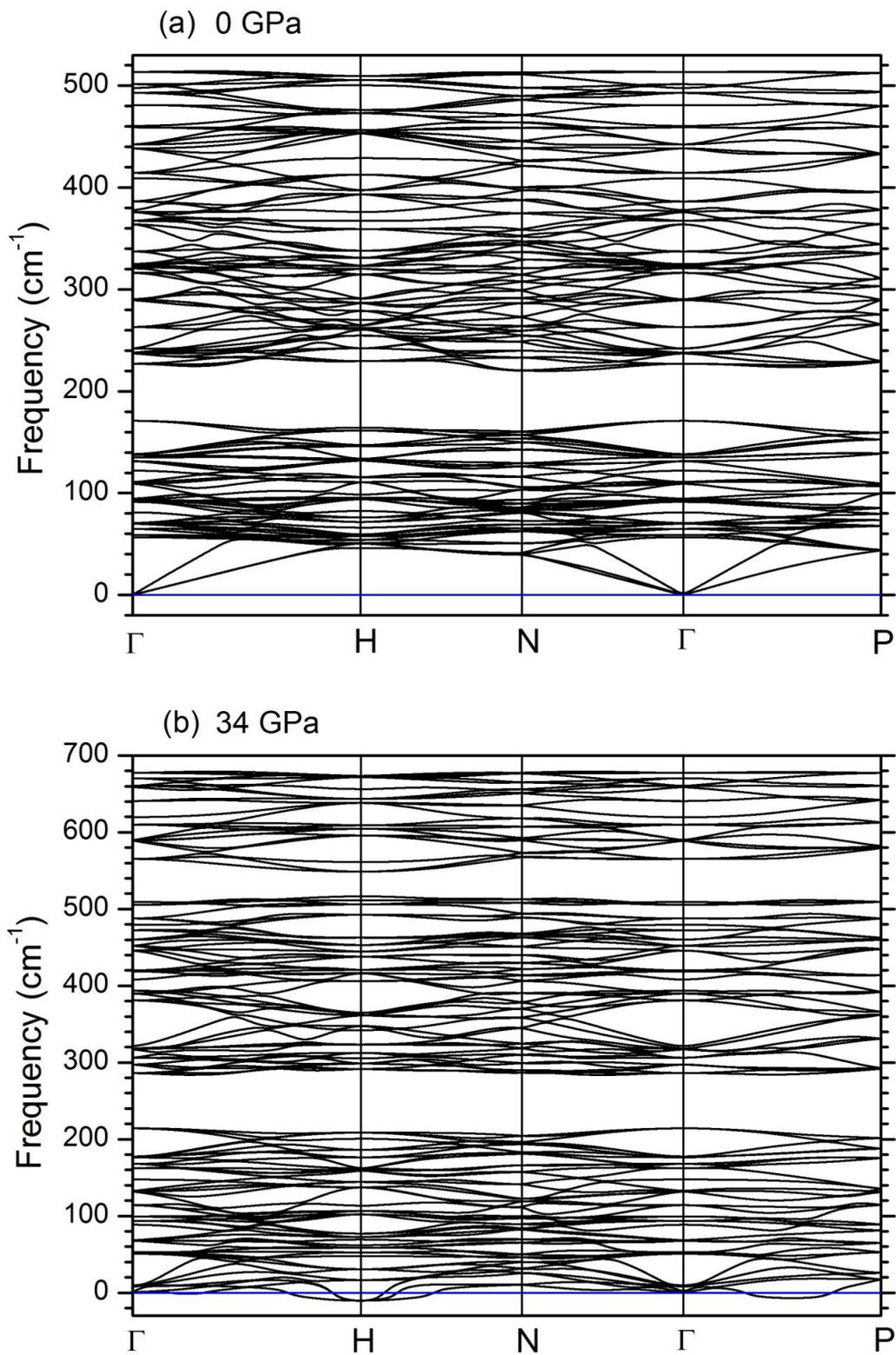

**Figure S1.** (Color online) Phonon dispersion curves in bixbyite-type $Tl_2O_3$ at 0 GPa (a) and 34 GPa (b).